\documentclass[prx,twocolumn,aps,superscriptaddress,nofootinbib,floatfix]{revtex4-2}

\usepackage{graphicx}
\usepackage{amsmath}
\usepackage{amssymb}
\usepackage{hyperref}

\begin{document}

\title{The machine use of human beings}

\author{Daniel W. Hook}
\email{daniel@digital-science.com}
\affiliation{Centre for Complexity Science, Imperial College London, London SW7 2AZ, United Kingdom}
\affiliation{Holtzbrinck, 6 Briset Street, London EC1M 5NR, United Kingdom}

\date{\today}

\begin{abstract}
Research results are prioritised by search engines, large language models and knowledge graphs chiefly through restatement rather than through the originating source. Although the open-access share of annual scholarly output has recently exceeded half, a substantial subscription corpus remains outside what machine readers may legitimately access under prevailing licences. A mechanism is described whereby subscription content is drawn into the open corpus through its citation and restatement in open-access articles---a process termed \emph{human mining}, by analogy with the text and data mining performed by machines. An optimistic upper bound on what may thereby be inferred by a machine reader confined to the open literature is modelled, and the completeness, latency and licence constraints of that bound are quantified. A majority of subscription articles that have ever been cited is found to be reachable through at least one open citation, and the median interval between a subscription article's publication and its first open citation is shown to have fallen from thirteen years for work of 1990 to one year for work of 2020. The advantage once conferred by direct subscription access to machine readers has thereby been largely eroded, chiefly as a consequence of the growth of open publishing rather than any change in how researchers cite.
\end{abstract}

\maketitle

The rise of generative artificial intelligence, and of large language models (LLMs) in particular, as a class of tools able to perform a range of functions in the research environment, now seems inexorable. Used responsibly, these tools augment our capabilities as researchers and quicken the progress of research. In reading, they summarise and recast complex material, helping us to grasp developments not only in our own fields but also in those further afield; in writing, they can sharpen the lucidity of even the most experienced author; and they assist, besides, with coding, automation, pattern recognition and much else. There are, of course, well-rehearsed downsides to the use of AI, which we need not revisit here; it suffices to say that LLMs have already demonstrated significant capability in research.

A substantial proportion of global research is publicly funded, and it is often argued that researchers have a corresponding responsibility to see that research outputs are put to the service of humanity. Even without the funding argument the same conclusion can be reached on grounds of social responsibility, or on several other reasonable bases. If this view is taken, and if LLMs learn not only what knowledge there is to build upon but also how to conduct research by consuming research, then it follows that they should be allowed to read as much of the scholarly literature as possible. We do not here assess the wisdom of that argument, but we note its logical force. Were LLMs to become capable of solving research problems at scale, and faster or more accurately or efficiently than humans, then it would be all the more important that they should have access to that literature.

Agreeing with the argument is, however, quite another thing from implementing it. The scholarly record was never designed to be machine readable at scale, and the landscape it presents to a machine reader is complex. It might be supposed that most of the open-access corpus is available for machine-learning applications while most of the subscription corpus is not. The second half of that statement is broadly correct; the first is less obviously so. Many open-access licences explicitly prohibit text and data mining for commercial purposes, and since a model may put the content it has read to commercial use in many settings, it is often difficult to argue that any particular use will not be commercial. That is only one of the frictions involved.

Given that LLMs may not formally reach around publisher paywalls and must licence subscription content directly where it is wanted, a reasonable question presents itself: given the existence of an open-access corpus that was itself built on the subscription corpus, do LLMs need to access what is behind the paywall at all?

The mechanism at the heart of this paper is simple. A researcher with subscription access reads an article and cites it in an open-access paper that they go on to publish. The citing author will typically restate, or imply, enough of the detail from the paper to make the citation intelligible; this is a legitimate and necessary feature of research communication. A search engine, a knowledge graph or a large language model that may lawfully read only the second article comes thereby to know some part of the content of the first. Each such citation can be thought of as a measurement, with the human author in the role of the measurement apparatus. We term this process \emph{human mining}, by analogy with the text and data mining performed by machines.

In 1950 Norbert Wiener argued that the danger of an age of automatic machines lay less in machines doing the work of people than in people being used as components in processes that they had neither designed nor could control~\cite{Wiener1950}. His title, \emph{The Human Use of Human Beings}, can be turned around, with some precision, to describe the mechanism just set out: the machine use of human beings, in which the researchers who read, synthesise and cite are the sensors through which the machines come to see the subscription literature.

For most of the history of scholarship, to know that a result was sound was to know where it had first been shown. The citation was designed around this convention, which we shall call \emph{proof of origin}: priority is adjudicated by it~\cite{Merton1957}, and the citation index was first conceived as a means of following references back to their source~\cite{Garfield1955}. While the literature was small and grew slowly, the most frequently cited statement of a result and its first statement were ordinarily the same document, and the distinction was of little practical consequence. The volume and pace of publication over recent decades have separated the two. A researcher can no longer read, or even find, most of what is relevant to their work, and the tools built to cope---web search, which ranks a document by the links pointing to it~\cite{Brin1998}, recommendation engines, and latterly LLMs, which encode a claim in proportion to how often it appears in their training text~\cite{kandpal2023longtail,mallen2023trust}---each resolve the choice between origin and restatement in favour of restatement. We call this \emph{proof of reuse}.

The substitution of the much-cited version of an idea for its first appearance is made by these systems by construction rather than by anyone's decision; it is, loosely, a ``microcanonical'' move, in which a temporal succession of developments running back to an originating source is replaced by a spatial ensemble of citing papers, each of which measures the same underlying fact and contributes to a graph from which that fact can in turn be inferred. The shift from a linear chain of developments to knowledge embodied in a graph is not only microcanonical in style but also closely related to the way that knowledge graphs are structured and used.

Proof of reuse has some merit, since a result restated many times by authors with no stake in its origin has been tested in a way that an unread original has not. Its failing is that it does not look for the originator, and that the route to being restated is uneven. Work is found in proportion to its accessibility, its author's reputation and position in a network, the clarity with which it was written and, increasingly, its authors' ability to pay for open publication. It would be naive to suppose that the older convention operated on a level field, but the new one has, as yet, no mechanism for correcting the unevenness, and the systems that apply to it carry no representation of where an idea began.

What follows constructs a model of human mining and uses it to place an optimistic upper bound on what a machine reader confined to the open-access literature can know about the content of the subscription corpus. We quantify the share of subscription articles that have crossed into the open through citation, the latency at which they do so, and the extent to which the picture is restricted by the licences of the open articles that do the citing.

\section*{A model for human-mined content}

What a machine may read is determined by licence. Text and data mining of subscription content is permitted in some jurisdictions for non-commercial research under statutory exception, and otherwise only by agreement, whereas open content may be mined if its licence terms allow. The subscription literature is therefore visible to machine readers legitimately only via inference using the open literature. In our model, human mining is measured subject to two assumptions: (i) that each article carries a single citable result, and (ii) that a single citation from an open article suffices to place that result in the open corpus. Both assumptions are generous, but they together form the basis of an optimistic bound on what the open literature holds of the subscription literature. Repeated citation may increase the fidelity and accuracy of what is transmitted, but we do not study that aspect here.

\textit{Data.}---We use Dimensions~\cite{Hook2018,Herzog2020} through Google BigQuery~\cite{Hook2021}, at a snapshot taken in September 2026, joined by DOI to Unpaywall~\cite{Piwowar2018}. We restrict the analysis to journal articles published from 1945 to 2025, on both the cited and the citing side. An article is open if Unpaywall classifies it as gold, hybrid or green. All other articles, including those classed as bronze, which are free to read on the publisher's site but carry no licence, are treated as subscription access, since reading at the publisher's discretion is not a basis on which a tool can be built. Green is assigned only where no free publisher copy exists, and the repository copy is the submitted version in 88--93\% of cases in every period since 1990, so that green is in practice a preprint flag. Citations are counted to the date of the snapshot, except in the year and licence analyses, which count citations made up to the end of 2025. The population comprises 124 million articles, of which 86.5 million are subscription access.

\begin{figure}
\includegraphics[width=\columnwidth]{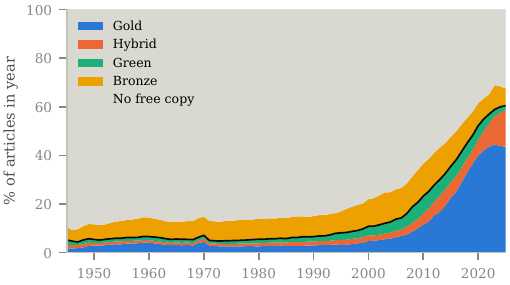}
\caption{Access route of journal articles by year of publication, as a share of each year's output. The black line is the open-access share (gold, hybrid and green); bronze is treated as subscription access.}
\label{fig:access_routes}
\end{figure}

Of the articles published since 1945, 30\% are open and 40\% are open or bronze (Fig.~\ref{fig:access_routes}). The open share of a single year's output passed one half in 2020 and reached 61\% in 2025, the greater part of it gold. Green reached 7\% of articles in 2010 and has since fallen below 2\% as preprinted articles have increasingly been published openly and so classified as gold or hybrid.

\begin{figure}
\includegraphics[width=\columnwidth]{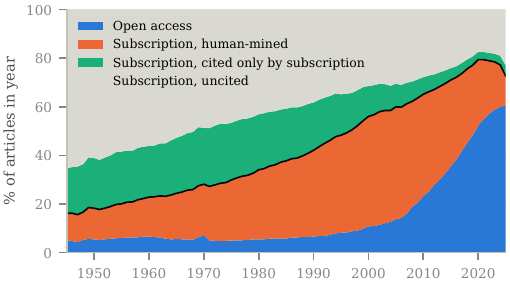}
\caption{Articles of each publication year divided into open access; subscription access and cited by at least one open article (human-mined); subscription access and cited only by subscription articles; and uncited. The black line is the share that is open or human-mined.}
\label{fig:mining_classes}
\end{figure}

Figure~\ref{fig:mining_classes} divides each year's articles into those that are open, those that are subscription and have been human-mined, those that are subscription and have been cited only by other subscription articles, and those that have not been cited at all. Over the whole period 44\% of subscription articles have been human-mined, and 61\% of all articles are either open or mined; for the articles of 2020 the figure is 79\% (52\% open and 27\% mined). An article that has never been cited, as is the case for 41\% of subscription articles, is all but invisible to a recommendation-driven or LLM-driven discovery approach, as only metadata are available to an algorithm or model. An article cited only from behind the subscription barrier is invisible to machine readers in particular, and it is this class that we use specifically to measure what the open corpus cannot see. This accounts for 14\% of subscription articles overall, a fall from 20\% of 1990 output to 3\% of that of 2020. Among subscription articles that have been cited at all, three quarters have been cited from the open literature.

\begin{figure}
\includegraphics[width=\columnwidth]{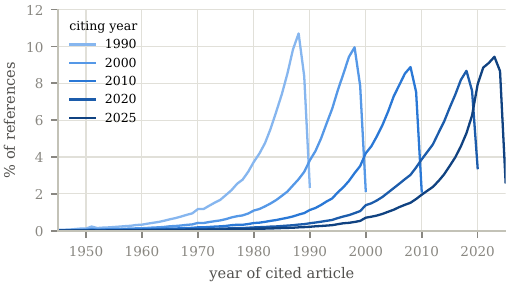}
\caption{Distribution of the publication years of the articles referenced by articles of five citing years, each normalised to 100\%.}
\label{fig:cited_year_profile}
\end{figure}

The speed at which results cross the black line in Fig.~\ref{fig:mining_classes} is governed by how rapidly the open literature makes citations. Figure~\ref{fig:cited_year_profile} shows the distribution of the publication years of the references made by all source publications in five selected years, and so answers the question of what constitutes the window of relevant research. Its form has been broadly stable since 1990: the peak lies two years before the source year, some 45\% of references are to the preceding five years and about 10\% to work more than twenty years old. The profile falls to half its peak within six to eight years, so that the window within which an article is likely to be cited is short. We note, however, the general trend toward a lower peak and greater spread, which we might conjecture is due to the increased use of search tools that highlight older articles.

\begin{figure}
\includegraphics[width=\columnwidth]{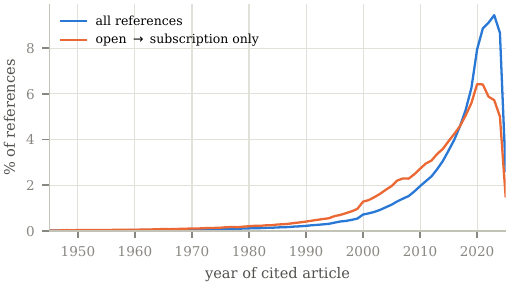}
\caption{For citing year 2025, the distribution of the publication years of all references (blue) and of references from open articles to subscription articles only (orange), each normalised to 100\%.}
\label{fig:open_vs_all}
\end{figure}

Figure~\ref{fig:open_vs_all} displays the difference between the resolving speed of using all citations and that of using only open citations to the subscription corpus, taking 2025 as the source year. The open-only measurement profile (orange) peaks lower and three years further back with a longer tail, and has a half-life of eight years against six years for the literature as a whole. Open citations reach the subscription literature less rapidly than citation in general but reach further into the past.

The same effect appears from the side of the cited article. For subscription articles published in 1990 the median interval between publication and first citation from any source was two years, and the median interval to first citation from an open article was thirteen. For the year of 2000 the second interval was seven years, for 2010 three and for 2020 one, while the first remained at one to two years throughout. A reader with access to the subscription literature of the 1990s knew, for a decade or more, results of which the open literature held no trace; for recent work that advantage has all but disappeared, as a consequence of the growth of open publishing rather than of any change in how researchers cite.

\textit{Licences.}---The open literature is not uniformly open to machines. The Creative Commons licences CC~BY and CC~BY-SA permit commercial reuse, the latter on condition that derivative works are shared on the same terms, which matters if a model or a knowledge graph is held to be a derivative. The non-commercial licences, CC~BY-NC and its variants, exclude commercial use, and the no-derivatives licences, CC~BY-ND and CC~BY-NC-ND, exclude derived works, which is the contested case for machine learning, since the extraction of facts is not generally held to create a derivative work whereas the reproduction of text may. Beyond these lie publisher-specific terms and much openly readable material, on preprint servers and in repositories, with no licence at all. Non-commercial and no-derivatives licences together have covered about a third of open articles in every year since 1990, while the share of open articles with no recorded licence has fallen from 47\% in 2000 to 11\% in 2025 and the share under CC~BY has risen from 21\% to 47\%.

\begin{figure}
\includegraphics[width=\columnwidth]{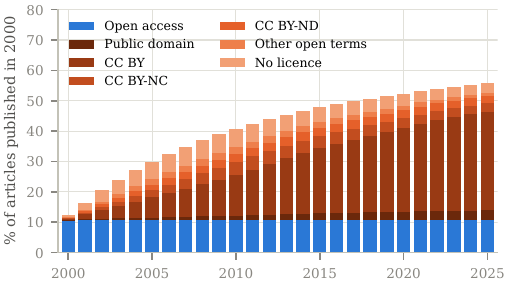}
\caption{Articles published in 2000: the open-access share (blue) and, above it, human-mined subscription articles, each attributed to the most permissive licence of any open article citing it by the year shown, from darkest (public domain) to lightest (no licence recorded).}
\label{fig:licence_decomposition}
\end{figure}

Figure~\ref{fig:licence_decomposition} decomposes the human-mined share of the articles of 2000 in this way. Of the whole year, 11\% is available directly as open access, each article under its own licence, and a further 45\% consists of subscription articles that had been human-mined by the end of 2025, which is to say about half of that year's subscription output. Of the year, 36\% can be reached through at least one public-domain or CC~BY citer, a further 2.7\% only through non-commercial licences, 2.5\% only through no-derivatives licences, and 4.1\% only through other terms or unlicensed copies. Between four fifths and nine tenths of the mined share is reachable through permissive citers for every year from 2000 to 2020. At this threshold the restriction costs little, because an article cited by many open articles is almost certain to have been cited by at least one under CC~BY. It costs more when several restatements are required: counting only permissive citers retains 0.79--0.88 of the subscription articles with at least one open citer, 0.58--0.72 of those with at least five and 0.48--0.60 of those with at least ten, depending on year. A tool confined to commercially reusable text therefore sees nearly as many subscription results as one that reads the whole open literature, but sees each of them through about half as many independent restatements, and the licence it relies upon is in any case the licence recorded in publisher metadata, which need not be the licence stated in the document.

\section*{An example}

\begin{figure}
\includegraphics[width=\columnwidth]{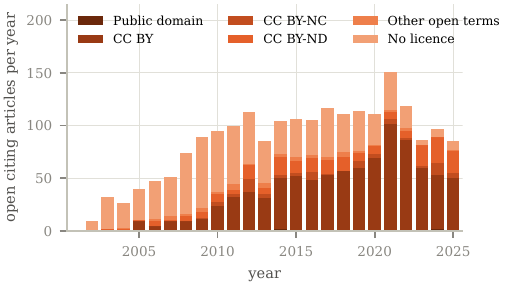}
\caption{Open-access articles citing Smith \textit{et al.}~\cite{Smith2000}, by year and by the licence of the citing article's best open copy, from darkest (public domain) to lightest (no licence recorded).}
\label{fig:smith_citations}
\end{figure}

The aggregate figures say how much of the subscription literature has an open projection via human mining, but not what it contains. To illustrate, we take Smith \textit{et al.}, who reported in 2000 the first composite medium with simultaneously negative permeability and permittivity~\cite{Smith2000}, realising a structure proposed by Pendry and colleagues~\cite{Pendry1999}; the version of record is free to read without a licence and is therefore, while generally available, classified as subscription access in our terms for machine-learning purposes. By the end of 2025 it had been cited by 2{,}066 open articles (Fig.~\ref{fig:smith_citations}). For its first four years it was reachable from the open literature almost entirely through unlicensed arXiv copies of subscription papers, which account for 63 of its first 68 open citers. The first CC~BY citations arrived in 2005, eight of them in a single issue of \emph{New Journal of Physics}; Unpaywall records these as CC~BY, although the PDFs themselves carry the publisher's copyright line and no licence statement.

The earliest citers name the result: ``By combining a 2D array of split-ring resonators (SRRs) with a 2D array of wires, Smith \textit{et al.} demonstrated for the first time the existence of left-handed materials''~\cite{Markos2002}; ``Their so-called meta-material consists of various inclusions of conducting rings and wires embedded within printed circuit boards''~\cite{McCall2002}; ``The first experiments indeed reported that the transmission of the LH samples was only $-20$~dB''~\cite{MarkosLosses2002}. From such sentences a reader of the open literature can recover what was built, what it demonstrated and that its interpretation was disputed, but not the frequency band, the geometry of the unit cell, or anything from the figures. The specific content of the result is carried by these early citations, which are the least likely to be permissively licensed, while the later, licensed citations more often refer to the paper as one of a class. Two of the 2005 citations also carry an error: Dewar credits the paper with demonstrating ``a negative index of refraction''~\cite{Dewar2005}, and Schurig and Smith place it among ``the experiments on negative refraction''~\cite{Schurig2005}, a demonstration that was in fact reported separately the following year~\cite{Shelby2001}. In a third it is attributed to the authors of a different Letter, although the reference resolves correctly~\cite{GarciaPomar2005}.

\section*{Discussion}

Human mining, as we have termed it, has carried a large proportion of the subscription literature into the open, and has done so faster, more completely and under more permissive terms with each successive year. What it carries is, however, selected. An article is mined via a citation measurement; the more measurements of the paper, the greater the fidelity of its content in the open sphere, up to a point. In many subjects it is not customary to quote passages or repeat narrative, but the key core results are often reasserted alongside a citation. While more moderate than the purely closed projection, the open projection of the subscription literature still favours the recent over the old, the popular over the obscure and the restatement over the origin, which are the preferences of the systems that read it. The originators of an idea are represented in that projection only to the extent that others chose to cite it, and those who did the citing---who are, in Wiener's sense, being used as the sensors through which the machines see---receive no credit for it.

The bound we have given is optimistic by construction. It treats a single citation as sufficient to carry a result into the open, and a measure of how often a citation in fact transmits the result, rather than the mere existence of the article, would require the citing sentences to be read at scale, which we have attempted here for only one paper. Nor is the analysis disaggregated by field or by language, although the practices that produce early open citation, preprinting above all, differ widely between disciplines.

There are two pressures on this boundary, and they pull in opposite directions. Many research funders carry an explicit mandate for the creation of wealth as well as a responsibility to support social and technological development, and a model that has read more of the literature is a better instrument for turning research into application, so that there is a public as well as a commercial interest in making as much of the record as possible available to such systems. Against this stands the recent experience of mathematicians. Within a single week of September 2026 a complete formalisation of Fermat's Last Theorem in Lean, some thirteen million lines produced by an AI system in eleven days, was built on the foundations of a funded project led by Kevin Buzzard to do the same~\cite{Buzzard2026,Castelvecchi2026FLT}, and a claimed resolution of the Navier--Stokes existence and smoothness problem was announced by OpenAI amid a dispute over whether the system had learned from the conversations of the mathematicians Buckmaster and Alp\"oge, who had been pursuing the same line of attack~\cite{Castelvecchi2026NS,Castelvecchi2026credit,Nature2026editorial}. In both cases the machine worked through what people had already done, and the question is how those people are to be credited, or displaced. What crosses the boundary, moreover, is a set of single results rather than the narrative and context in which they were argued---the method, the caveat, the approach that failed---and much of that remains behind subscription barriers, where it may be counted either as a protection for those who produced it or as an obstacle to those who would build upon it, according to which of the two pressures one feels more keenly.

\section*{Data availability}

The SQL, the exported result tables and the code used to produce the figures are available on Figshare: \url{https://doi.org/10.6084/m9.figshare.34065672}. Tables derived from Unpaywall require a separate copy of the Unpaywall database to reproduce.

\section*{Conflict of interest}

The author is an employee of Digital Science, the owner and operator of Dimensions.

\section*{AI usage}

Claude Sonnet 5.5 was used to improve and check SQL and python code used in creating the figures and data used in this paper.  Claude Sonnet 5.5 assisted by copy-editing this manuscript.  DWH takes responsibility for the factual accuracy of all analyses and commentary herein.

\bibliographystyle{apsrev4-2}
\bibliography{refs_prl}

\begin{thebibliography}{24}%
\makeatletter
\providecommand \@ifxundefined [1]{%
 \@ifx{#1\undefined}
}%
\providecommand \@ifnum [1]{%
 \ifnum #1\expandafter \@firstoftwo
 \else \expandafter \@secondoftwo
 \fi
}%
\providecommand \@ifx [1]{%
 \ifx #1\expandafter \@firstoftwo
 \else \expandafter \@secondoftwo
 \fi
}%
\providecommand \natexlab [1]{#1}%
\providecommand \enquote  [1]{``#1''}%
\providecommand \bibnamefont  [1]{#1}%
\providecommand \bibfnamefont [1]{#1}%
\providecommand \citenamefont [1]{#1}%
\providecommand \href@noop [0]{\@secondoftwo}%
\providecommand \href [0]{\begingroup \@sanitize@url \@href}%
\providecommand \@href[1]{\@@startlink{#1}\@@href}%
\providecommand \@@href[1]{\endgroup#1\@@endlink}%
\providecommand \@sanitize@url [0]{\catcode `\\12\catcode `\$12\catcode `\&12\catcode `\#12\catcode `\^12\catcode `\_12\catcode `\%12\relax}%
\providecommand \@@startlink[1]{}%
\providecommand \@@endlink[0]{}%
\providecommand \url  [0]{\begingroup\@sanitize@url \@url }%
\providecommand \@url [1]{\endgroup\@href {#1}{\urlprefix }}%
\providecommand \urlprefix  [0]{URL }%
\providecommand \Eprint [0]{\href }%
\providecommand \doibase [0]{https://doi.org/}%
\providecommand \selectlanguage [0]{\@gobble}%
\providecommand \bibinfo  [0]{\@secondoftwo}%
\providecommand \bibfield  [0]{\@secondoftwo}%
\providecommand \translation [1]{[#1]}%
\providecommand \BibitemOpen [0]{}%
\providecommand \bibitemStop [0]{}%
\providecommand \bibitemNoStop [0]{.\EOS\space}%
\providecommand \EOS [0]{\spacefactor3000\relax}%
\providecommand \BibitemShut  [1]{\csname bibitem#1\endcsname}%
\let\auto@bib@innerbib\@empty
%</preamble>
\bibitem [{\citenamefont {Wiener}(1950)}]{Wiener1950}%
  \BibitemOpen
  \bibfield  {author} {\bibinfo {author} {\bibfnamefont {N.}~\bibnamefont {Wiener}},\ }\href@noop {} {\emph {\bibinfo {title} {The Human Use of Human Beings: Cybernetics and Society}}}\ (\bibinfo  {publisher} {Houghton Mifflin},\ \bibinfo {address} {Boston},\ \bibinfo {year} {1950})\BibitemShut {NoStop}%
\bibitem [{\citenamefont {Merton}(1957)}]{Merton1957}%
  \BibitemOpen
  \bibfield  {author} {\bibinfo {author} {\bibfnamefont {R.~K.}\ \bibnamefont {Merton}},\ }\href {https://doi.org/10.2307/2089193} {\bibfield  {journal} {\bibinfo  {journal} {American Sociological Review}\ }\textbf {\bibinfo {volume} {22}},\ \bibinfo {pages} {635} (\bibinfo {year} {1957})}\BibitemShut {NoStop}%
\bibitem [{\citenamefont {Garfield}(1955)}]{Garfield1955}%
  \BibitemOpen
  \bibfield  {author} {\bibinfo {author} {\bibfnamefont {E.}~\bibnamefont {Garfield}},\ }\href {https://doi.org/10.1126/science.122.3159.108} {\bibfield  {journal} {\bibinfo  {journal} {Science}\ }\textbf {\bibinfo {volume} {122}},\ \bibinfo {pages} {108} (\bibinfo {year} {1955})}\BibitemShut {NoStop}%
\bibitem [{\citenamefont {Brin}\ and\ \citenamefont {Page}(1998)}]{Brin1998}%
  \BibitemOpen
  \bibfield  {author} {\bibinfo {author} {\bibfnamefont {S.}~\bibnamefont {Brin}}\ and\ \bibinfo {author} {\bibfnamefont {L.}~\bibnamefont {Page}},\ }\href {https://doi.org/10.1016/S0169-7552(98)00110-X} {\bibfield  {journal} {\bibinfo  {journal} {Computer Networks and ISDN Systems}\ }\textbf {\bibinfo {volume} {30}},\ \bibinfo {pages} {107} (\bibinfo {year} {1998})}\BibitemShut {NoStop}%
\bibitem [{\citenamefont {Kandpal}\ \emph {et~al.}(2023)\citenamefont {Kandpal}, \citenamefont {Deng}, \citenamefont {Roberts}, \citenamefont {Wallace},\ and\ \citenamefont {Raffel}}]{kandpal2023longtail}%
  \BibitemOpen
  \bibfield  {author} {\bibinfo {author} {\bibfnamefont {N.}~\bibnamefont {Kandpal}}, \bibinfo {author} {\bibfnamefont {H.}~\bibnamefont {Deng}}, \bibinfo {author} {\bibfnamefont {A.}~\bibnamefont {Roberts}}, \bibinfo {author} {\bibfnamefont {E.}~\bibnamefont {Wallace}},\ and\ \bibinfo {author} {\bibfnamefont {C.}~\bibnamefont {Raffel}},\ }in\ \href@noop {} {\emph {\bibinfo {booktitle} {Proceedings of the 40th International Conference on Machine Learning}}},\ \bibinfo {series} {Proceedings of Machine Learning Research}, Vol.\ \bibinfo {volume} {202}\ (\bibinfo  {publisher} {PMLR},\ \bibinfo {year} {2023})\ pp.\ \bibinfo {pages} {15696--15707},\ \Eprint {https://arxiv.org/abs/2211.08411} {arXiv:2211.08411 [cs.CL]} \BibitemShut {NoStop}%
\bibitem [{\citenamefont {Mallen}\ \emph {et~al.}(2023)\citenamefont {Mallen}, \citenamefont {Asai}, \citenamefont {Zhong}, \citenamefont {Das}, \citenamefont {Khashabi},\ and\ \citenamefont {Hajishirzi}}]{mallen2023trust}%
  \BibitemOpen
  \bibfield  {author} {\bibinfo {author} {\bibfnamefont {A.}~\bibnamefont {Mallen}}, \bibinfo {author} {\bibfnamefont {A.}~\bibnamefont {Asai}}, \bibinfo {author} {\bibfnamefont {V.}~\bibnamefont {Zhong}}, \bibinfo {author} {\bibfnamefont {R.}~\bibnamefont {Das}}, \bibinfo {author} {\bibfnamefont {D.}~\bibnamefont {Khashabi}},\ and\ \bibinfo {author} {\bibfnamefont {H.}~\bibnamefont {Hajishirzi}},\ }in\ \href {https://doi.org/10.18653/v1/2023.acl-long.546} {\emph {\bibinfo {booktitle} {Proceedings of the 61st Annual Meeting of the Association for Computational Linguistics (Volume 1: Long Papers)}}},\ \bibinfo {editor} {edited by\ \bibinfo {editor} {\bibfnamefont {A.}~\bibnamefont {Rogers}}, \bibinfo {editor} {\bibfnamefont {J.}~\bibnamefont {Boyd-Graber}},\ and\ \bibinfo {editor} {\bibfnamefont {N.}~\bibnamefont {Okazaki}}}\ (\bibinfo  {publisher} {Association for Computational Linguistics},\ \bibinfo {address} {Toronto, Canada},\ \bibinfo {year} {2023})\ pp.\ \bibinfo {pages} {9802--9822},\ \Eprint
  {https://arxiv.org/abs/2212.10511} {arXiv:2212.10511 [cs.CL]} \BibitemShut {NoStop}%
\bibitem [{\citenamefont {Hook}\ \emph {et~al.}(2018)\citenamefont {Hook}, \citenamefont {Porter},\ and\ \citenamefont {Herzog}}]{Hook2018}%
  \BibitemOpen
  \bibfield  {author} {\bibinfo {author} {\bibfnamefont {D.~W.}\ \bibnamefont {Hook}}, \bibinfo {author} {\bibfnamefont {S.~J.}\ \bibnamefont {Porter}},\ and\ \bibinfo {author} {\bibfnamefont {C.}~\bibnamefont {Herzog}},\ }\href {https://doi.org/10.3389/frma.2018.00023} {\bibfield  {journal} {\bibinfo  {journal} {Frontiers in Research Metrics and Analytics}\ }\textbf {\bibinfo {volume} {3}},\ \bibinfo {pages} {23} (\bibinfo {year} {2018})}\BibitemShut {NoStop}%
\bibitem [{\citenamefont {Herzog}\ \emph {et~al.}(2020)\citenamefont {Herzog}, \citenamefont {Hook},\ and\ \citenamefont {Konkiel}}]{Herzog2020}%
  \BibitemOpen
  \bibfield  {author} {\bibinfo {author} {\bibfnamefont {C.}~\bibnamefont {Herzog}}, \bibinfo {author} {\bibfnamefont {D.}~\bibnamefont {Hook}},\ and\ \bibinfo {author} {\bibfnamefont {S.}~\bibnamefont {Konkiel}},\ }\href {https://doi.org/10.1162/qss_a_00020} {\bibfield  {journal} {\bibinfo  {journal} {Quantitative Science Studies}\ }\textbf {\bibinfo {volume} {1}},\ \bibinfo {pages} {387} (\bibinfo {year} {2020})}\BibitemShut {NoStop}%
\bibitem [{\citenamefont {Hook}\ and\ \citenamefont {Porter}(2021)}]{Hook2021}%
  \BibitemOpen
  \bibfield  {author} {\bibinfo {author} {\bibfnamefont {D.~W.}\ \bibnamefont {Hook}}\ and\ \bibinfo {author} {\bibfnamefont {S.~J.}\ \bibnamefont {Porter}},\ }\href {https://doi.org/10.3389/frma.2021.656233} {\bibfield  {journal} {\bibinfo  {journal} {Frontiers in Research Metrics and Analytics}\ }\textbf {\bibinfo {volume} {6}},\ \bibinfo {pages} {656233} (\bibinfo {year} {2021})}\BibitemShut {NoStop}%
\bibitem [{\citenamefont {Piwowar}\ \emph {et~al.}(2018)\citenamefont {Piwowar}, \citenamefont {Priem}, \citenamefont {Larivi{\`e}re}, \citenamefont {Alperin}, \citenamefont {Matthias}, \citenamefont {Norlander}, \citenamefont {Farley}, \citenamefont {West},\ and\ \citenamefont {Haustein}}]{Piwowar2018}%
  \BibitemOpen
  \bibfield  {author} {\bibinfo {author} {\bibfnamefont {H.}~\bibnamefont {Piwowar}}, \bibinfo {author} {\bibfnamefont {J.}~\bibnamefont {Priem}}, \bibinfo {author} {\bibfnamefont {V.}~\bibnamefont {Larivi{\`e}re}}, \bibinfo {author} {\bibfnamefont {J.~P.}\ \bibnamefont {Alperin}}, \bibinfo {author} {\bibfnamefont {L.}~\bibnamefont {Matthias}}, \bibinfo {author} {\bibfnamefont {B.}~\bibnamefont {Norlander}}, \bibinfo {author} {\bibfnamefont {A.}~\bibnamefont {Farley}}, \bibinfo {author} {\bibfnamefont {J.}~\bibnamefont {West}},\ and\ \bibinfo {author} {\bibfnamefont {S.}~\bibnamefont {Haustein}},\ }\href {https://doi.org/10.7717/peerj.4375} {\bibfield  {journal} {\bibinfo  {journal} {PeerJ}\ }\textbf {\bibinfo {volume} {6}},\ \bibinfo {pages} {e4375} (\bibinfo {year} {2018})}\BibitemShut {NoStop}%
\bibitem [{\citenamefont {Smith}\ \emph {et~al.}(2000)\citenamefont {Smith}, \citenamefont {Padilla}, \citenamefont {Vier}, \citenamefont {Nemat-Nasser},\ and\ \citenamefont {Schultz}}]{Smith2000}%
  \BibitemOpen
  \bibfield  {author} {\bibinfo {author} {\bibfnamefont {D.~R.}\ \bibnamefont {Smith}}, \bibinfo {author} {\bibfnamefont {W.~J.}\ \bibnamefont {Padilla}}, \bibinfo {author} {\bibfnamefont {D.~C.}\ \bibnamefont {Vier}}, \bibinfo {author} {\bibfnamefont {S.~C.}\ \bibnamefont {Nemat-Nasser}},\ and\ \bibinfo {author} {\bibfnamefont {S.}~\bibnamefont {Schultz}},\ }\href {https://doi.org/10.1103/PhysRevLett.84.4184} {\bibfield  {journal} {\bibinfo  {journal} {Physical Review Letters}\ }\textbf {\bibinfo {volume} {84}},\ \bibinfo {pages} {4184} (\bibinfo {year} {2000})}\BibitemShut {NoStop}%
\bibitem [{\citenamefont {Pendry}\ \emph {et~al.}(1999)\citenamefont {Pendry}, \citenamefont {Holden}, \citenamefont {Robbins},\ and\ \citenamefont {Stewart}}]{Pendry1999}%
  \BibitemOpen
  \bibfield  {author} {\bibinfo {author} {\bibfnamefont {J.~B.}\ \bibnamefont {Pendry}}, \bibinfo {author} {\bibfnamefont {A.~J.}\ \bibnamefont {Holden}}, \bibinfo {author} {\bibfnamefont {D.~J.}\ \bibnamefont {Robbins}},\ and\ \bibinfo {author} {\bibfnamefont {W.~J.}\ \bibnamefont {Stewart}},\ }\href {https://doi.org/10.1109/22.798002} {\bibfield  {journal} {\bibinfo  {journal} {IEEE Transactions on Microwave Theory and Techniques}\ }\textbf {\bibinfo {volume} {47}},\ \bibinfo {pages} {2075} (\bibinfo {year} {1999})}\BibitemShut {NoStop}%
\bibitem [{\citenamefont {Marko{\v{s}}}\ and\ \citenamefont {Soukoulis}(2001)}]{Markos2002}%
  \BibitemOpen
  \bibfield  {author} {\bibinfo {author} {\bibfnamefont {P.}~\bibnamefont {Marko{\v{s}}}}\ and\ \bibinfo {author} {\bibfnamefont {C.~M.}\ \bibnamefont {Soukoulis}},\ }\href {https://doi.org/10.1103/PhysRevB.65.033401} {\bibfield  {journal} {\bibinfo  {journal} {Physical Review B}\ }\textbf {\bibinfo {volume} {65}},\ \bibinfo {pages} {033401} (\bibinfo {year} {2001})}\BibitemShut {NoStop}%
\bibitem [{\citenamefont {McCall}\ \emph {et~al.}(2002)\citenamefont {McCall}, \citenamefont {Lakhtakia},\ and\ \citenamefont {Weiglhofer}}]{McCall2002}%
  \BibitemOpen
  \bibfield  {author} {\bibinfo {author} {\bibfnamefont {M.~W.}\ \bibnamefont {McCall}}, \bibinfo {author} {\bibfnamefont {A.}~\bibnamefont {Lakhtakia}},\ and\ \bibinfo {author} {\bibfnamefont {W.~S.}\ \bibnamefont {Weiglhofer}},\ }\href {https://doi.org/10.1088/0143-0807/23/3/314} {\bibfield  {journal} {\bibinfo  {journal} {European Journal of Physics}\ }\textbf {\bibinfo {volume} {23}},\ \bibinfo {pages} {353} (\bibinfo {year} {2002})}\BibitemShut {NoStop}%
\bibitem [{\citenamefont {Marko{\v{s}}}\ \emph {et~al.}(2002)\citenamefont {Marko{\v{s}}}, \citenamefont {Rousochatzakis},\ and\ \citenamefont {Soukoulis}}]{MarkosLosses2002}%
  \BibitemOpen
  \bibfield  {author} {\bibinfo {author} {\bibfnamefont {P.}~\bibnamefont {Marko{\v{s}}}}, \bibinfo {author} {\bibfnamefont {I.}~\bibnamefont {Rousochatzakis}},\ and\ \bibinfo {author} {\bibfnamefont {C.~M.}\ \bibnamefont {Soukoulis}},\ }\href {https://doi.org/10.1103/PhysRevE.66.045601} {\bibfield  {journal} {\bibinfo  {journal} {Physical Review E}\ }\textbf {\bibinfo {volume} {66}},\ \bibinfo {pages} {045601} (\bibinfo {year} {2002})}\BibitemShut {NoStop}%
\bibitem [{\citenamefont {Dewar}(2005)}]{Dewar2005}%
  \BibitemOpen
  \bibfield  {author} {\bibinfo {author} {\bibfnamefont {G.}~\bibnamefont {Dewar}},\ }\href {https://doi.org/10.1088/1367-2630/7/1/161} {\bibfield  {journal} {\bibinfo  {journal} {New Journal of Physics}\ }\textbf {\bibinfo {volume} {7}},\ \bibinfo {pages} {161} (\bibinfo {year} {2005})}\BibitemShut {NoStop}%
\bibitem [{\citenamefont {Schurig}\ and\ \citenamefont {Smith}(2005)}]{Schurig2005}%
  \BibitemOpen
  \bibfield  {author} {\bibinfo {author} {\bibfnamefont {D.}~\bibnamefont {Schurig}}\ and\ \bibinfo {author} {\bibfnamefont {D.~R.}\ \bibnamefont {Smith}},\ }\href {https://doi.org/10.1088/1367-2630/7/1/162} {\bibfield  {journal} {\bibinfo  {journal} {New Journal of Physics}\ }\textbf {\bibinfo {volume} {7}},\ \bibinfo {pages} {162} (\bibinfo {year} {2005})}\BibitemShut {NoStop}%
\bibitem [{\citenamefont {Shelby}\ \emph {et~al.}(2001)\citenamefont {Shelby}, \citenamefont {Smith},\ and\ \citenamefont {Schultz}}]{Shelby2001}%
  \BibitemOpen
  \bibfield  {author} {\bibinfo {author} {\bibfnamefont {R.~A.}\ \bibnamefont {Shelby}}, \bibinfo {author} {\bibfnamefont {D.~R.}\ \bibnamefont {Smith}},\ and\ \bibinfo {author} {\bibfnamefont {S.}~\bibnamefont {Schultz}},\ }\href {https://doi.org/10.1126/science.1058847} {\bibfield  {journal} {\bibinfo  {journal} {Science}\ }\textbf {\bibinfo {volume} {292}},\ \bibinfo {pages} {77} (\bibinfo {year} {2001})}\BibitemShut {NoStop}%
\bibitem [{\citenamefont {Garcia-Pomar}\ and\ \citenamefont {Nieto-Vesperinas}(2005)}]{GarciaPomar2005}%
  \BibitemOpen
  \bibfield  {author} {\bibinfo {author} {\bibfnamefont {J.~L.}\ \bibnamefont {Garcia-Pomar}}\ and\ \bibinfo {author} {\bibfnamefont {M.}~\bibnamefont {Nieto-Vesperinas}},\ }\href {https://doi.org/10.1088/1367-2630/7/1/160} {\bibfield  {journal} {\bibinfo  {journal} {New Journal of Physics}\ }\textbf {\bibinfo {volume} {7}},\ \bibinfo {pages} {160} (\bibinfo {year} {2005})}\BibitemShut {NoStop}%
\bibitem [{\citenamefont {Buzzard}(2026)}]{Buzzard2026}%
  \BibitemOpen
  \bibfield  {author} {\bibinfo {author} {\bibfnamefont {K.}~\bibnamefont {Buzzard}},\ }\href {https://xenaproject.wordpress.com/2026/09/04/flt-anthropic-has-beaten-me-to-it/} {\bibinfo {title} {{FLT}: {A}nthropic has beaten me to it}},\ \bibinfo {howpublished} {Xena Project blog, 4 September 2026} (\bibinfo {year} {2026})\BibitemShut {NoStop}%
\bibitem [{\citenamefont {Castelvecchi}(2026{\natexlab{a}})}]{Castelvecchi2026FLT}%
  \BibitemOpen
  \bibfield  {author} {\bibinfo {author} {\bibfnamefont {D.}~\bibnamefont {Castelvecchi}},\ }\bibfield  {journal} {\bibinfo  {journal} {Nature}\ }\href {https://doi.org/10.1038/d41586-026-02822-9} {10.1038/d41586-026-02822-9} (\bibinfo {year} {2026}{\natexlab{a}}),\ \bibinfo {note} {news, 7 September 2026}\BibitemShut {NoStop}%
\bibitem [{\citenamefont {Castelvecchi}(2026{\natexlab{b}})}]{Castelvecchi2026NS}%
  \BibitemOpen
  \bibfield  {author} {\bibinfo {author} {\bibfnamefont {D.}~\bibnamefont {Castelvecchi}},\ }\bibfield  {journal} {\bibinfo  {journal} {Nature}\ }\href {https://doi.org/10.1038/d41586-026-02842-5} {10.1038/d41586-026-02842-5} (\bibinfo {year} {2026}{\natexlab{b}}),\ \bibinfo {note} {news, 8 September 2026}\BibitemShut {NoStop}%
\bibitem [{\citenamefont {Castelvecchi}(2026{\natexlab{c}})}]{Castelvecchi2026credit}%
  \BibitemOpen
  \bibfield  {author} {\bibinfo {author} {\bibfnamefont {D.}~\bibnamefont {Castelvecchi}},\ }\bibfield  {journal} {\bibinfo  {journal} {Nature}\ }\href {https://doi.org/10.1038/d41586-026-02910-w} {10.1038/d41586-026-02910-w} (\bibinfo {year} {2026}{\natexlab{c}}),\ \bibinfo {note} {news, 17 September 2026}\BibitemShut {NoStop}%
\bibitem [{\citenamefont {{Nature}}(2026)}]{Nature2026editorial}%
  \BibitemOpen
  \bibfield  {author} {\bibinfo {author} {\bibnamefont {{Nature}}},\ }\href {https://doi.org/10.1038/d41586-026-02886-7} {\bibfield  {journal} {\bibinfo  {journal} {Nature}\ }\textbf {\bibinfo {volume} {657}},\ \bibinfo {pages} {573} (\bibinfo {year} {2026})},\ \bibinfo {note} {editorial, 16 September 2026}\BibitemShut {NoStop}%
\end{thebibliography}%

\end{document}